\documentclass[aps,pra,amsfonts,amssymb,twocolumn,groupedaddress,showpacs,floatfix]{revtex4}
\usepackage{graphicx}
\usepackage{graphics}
\usepackage{amsmath}

\begin{document}

\title{Mediated Homogenization}

\author{Daniel Burgarth$^{1}$ and Vittorio Giovannetti$^{2}$}

\affiliation{
$^{1}$Computer Science Department, ETH Z\"urich, CH-8092 Z\"urich, Switzerland\\
$^{2}$NEST-CNR-INFM \& Scuola Normale Superiore, piazza dei Cavalieri 7, I-56126 Pisa, Italy
}

\begin{abstract}

Homogenization protocols model the quantum mechanical evolution
of a system to a fixed state independently
from its initial configuration by repeatedly 
coupling it with a collection of identical ancillas.
Here we analyze these protocols within the formalism of ``relaxing'' channels
providing an easy to check sufficient condition for homogenization.
In this context we describe mediated homogenization schemes where a network of connected qudits
relaxes to a fixed state by only
partially interacting  with a bath.
We also study configurations which allow us to
introduce entanglement among the elements of the network. Finally we analyze 
the effect of having competitive configurations with two different baths and we prove
the convergence to dynamical equilibrium for Heisenberg chains.
\end{abstract}

\maketitle

\section{Introduction}\label{s:theset}

Homogenization protocols have been extensively studied
in recent years as a powerful model for the equilibration of a quantum mechanical system  interacting
with a large bath~\cite{HOMOGENIZATION1,HOMOGENIZATION2,Scarani,conti,ckw,qudit}. 
In these schemes one considers 
a collision-like  coupling of the system with a collection of ancillary systems
that have been prepared in identical states. 
This corresponds to a Markovian approximation in a discrete
dynamical evolution. Compared to typical quantum Markov equations the advantage of this model is that it allows one
to concentrate on the effective unitaries and completely positive maps rather than the underlying Hamiltonians and Lindblad generators~\cite{Breuer}.
By ``homogenization'' one means that the system state converges to a state that is the same as the ancilla states. This was demonstrated for a class of qudit systems in~\cite{HOMOGENIZATION1,HOMOGENIZATION2,qudit}. The bath-system entanglement was studied in~\cite{ckw}, and a continuous-time model (quantum master equation) was derived from the discrete model in~\cite{conti}.
Furthermore, the emergence of irreversibility was investigated in~\cite{Scarani}.

An important aspect of quantum homogenization is that is as a stable method of driving a system into some fixed state, independent of its
initial state. In this context, we will also refer to the bath as a ``controller'' system and its state as a ``controller state''.
Hence apart from its fundamental role of studying quantum convergence, homogenization has possible applications for quantum cloning~\cite{HOMOGENIZATION2}, for the hiding of quantum information~\cite{Winter,HOMOGENIZATION2} and
spin chain quantum communication~\cite{MEMORYSWAP}.

%%%%%%%%%%%%%%%%%%%%%%%%%%%%%%%%%%%%%%%%%%%%%%%%%%%%
\begin{figure}[t]
\includegraphics[width=0.7\columnwidth]{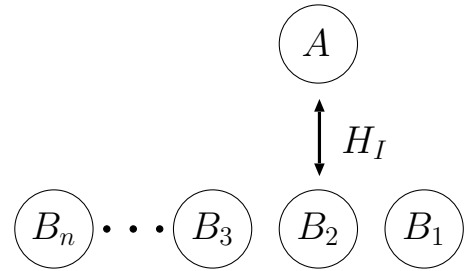}
\caption{\label{setup}Setup of a standard homogenization protocol:
the controlled system $A$ interact with a collection of controller
systems $B$ which have been initialized into the same input state
$\omega_B$. Homogenization takes place when the final state of $A$ is driven into the same state
 $\omega_A$ of the controllers in the limit of infinitely many
couplings with the $B$s.}
\end{figure}
%%%%%%%%%%%%%%%%%%%%%%%%%%%%%%%%%%%%%%%%%%%%%%%%%%%%%%%%%%
The prototypical homogenization scenario~\cite{HOMOGENIZATION1,HOMOGENIZATION2}
 is described in Fig.~\ref{setup}.
It is composed of two parts: a system $A$ with
an always on  Hamiltonian $H_A$, and a large ensemble of identical 
controller systems $B_1, B_2, \cdots,B_n$.
The latter are prepared in the same  initial
state $\omega_B$ and 
are assumed to have 
no independent free evolution.
The system $A$ is coupled in sequential order with
 each one of the $B$s 
through a series of identical stepwise 
interactions described by 
the Hamiltonian $H_I$.
In this setting the evolution of the system $A$ 
is described by the successive application of
the completely positive (CP) map
\begin{eqnarray}
{\cal E}(\rho_{A})\equiv \mbox{Tr}_{B}\left[ 
U (\rho_{A}\otimes \omega_B) U^{\dag}\right]  \;, \label{mappa}
\end{eqnarray}
with $U\equiv {\exp}[ -i (H_A + H_I)t ]$ and $t>0$ 
being the time interval associated with a single $A$-$B$ coupling.
After the interaction with $n$ controllers 
the state of $A$ becomes
\begin{eqnarray}
\rho_A^{(n)} = \underbrace{{\cal E} \circ {\cal E} \circ
\cdots \circ {\cal E}}_{\mbox{$n$ times}}(\rho_A) \equiv {\cal E}^n(\rho_{A}) \;.
\label{seq}
\end{eqnarray}
We are interested in the behavior of the
sequence~(\ref{seq}) for large $n$:
in the case where the system $A$  and the controllers $B_k$ are identical, $H_A=0$ and $H_I$ is a swap Hamiltonian it was
shown~\cite{HOMOGENIZATION1,HOMOGENIZATION2,qudit} that the state of $A$ asymptotically converges to the state $\omega_B$ of the controllers,
independently from the initial state $\rho_A$.

In the above, homogenization also gives rise to thermalization~\cite{Terhal} - if the bath is initialized in Gibbs states, then the
system converges to a Gibbs state. However it is an open question if this still holds in a situation where system and bath particles have different dimensionality~\cite{Scarani}. Moreover, in~\cite{HOMOGENIZATION1,HOMOGENIZATION2,Scarani,conti,ckw,qudit} the bath is modeled to interact with the \emph{whole} system whereas an interaction with a subsystem (such as the \emph{surface}) seems more plausible.

A first generalization towards this direction
was observed by the Authors of the present paper
when studying the propagation
of quantum information along spin chain communication 
channels~\cite{MEMORYSWAP}.
In that case $A$ 
represented a collection of $N$ coupled qubits,  
while the Hamiltonians $H_I$  
implemented  a sequence of 
strong instantaneous swaps among the last element of the chain
 and a collection of  controller qubits (the $B$s).
By assuming the $B$s to be prepared into the spin down state $|0\rangle_B$
we showed that in the limit of large $n$, 
any initial state of $A$ will be coherently
transferred into the $B$s while the chain will be mapped into the all spin
down configuration $|00 \cdots 0 \rangle$ (the only requirement
being a non trivial connection among the qubits of $A$).
%%%%%%%%%%%%%%%%%%%%%%%%%%%%%%%%%%%%%%%%%%%%%%%%%%%%
\begin{figure}[t]
\includegraphics[width=0.7\columnwidth]{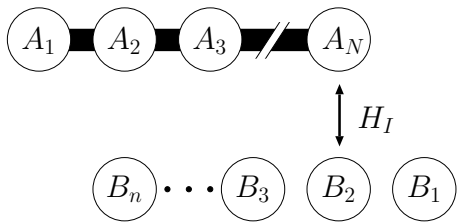}
\caption{\label{figura2}Generalized homogenization protocol:
in this case the controlled system $A$ is a composite one (e.g. a network
of coupled spins). Only a proper subset of the network interacts directly
with the controllers $B$ (as in the case of Fig.~\ref{setup}, the $B_k$ are assumed to be prepared in the
same initial states.) }
\end{figure}
%%%%%%%%%%%%%%%%%%%%%%%%%%%%%%%%%%%%%%%%%%%%%%%%%%%%%%%%%%
Concerning Ref.~\cite{MEMORYSWAP} 
it is worth stressing 
that in contrast to Refs.~\cite{HOMOGENIZATION1,HOMOGENIZATION2,Scarani,conti,ckw,qudit} $A$ and the controllers are quite distinct objects (namely $A$ is
a network of coupled qubits while each of the $B_1 \cdots B_n$
is just a single qubit). Moreover only a proper subset of $A$ (specifically
the $N$-th element of the network) interacts  directly with $B$: the remaining
qubits are only affected by the controllers  through the 
free Hamiltonian $H_A$ of the network (see Fig.~\ref{figura2}).
The possibility of preparing the ancilla state $|0\rangle_B$ 
into each one of the spins of the network
is therefore  a remarkable feature of the system which requires
some further investigation. We call it {\em mediated homogenization process}.

In this paper we tackle this issue 
by analyzing spin networks which show similar
properties. In particular in Sec.~\ref{s:heis}  we present a first example of a mediated  
homogenization process which allows one to transfer on the network element
{\em any} input state $\omega_B$ of the controllers.  
Before doing so however we introduce a general convergence criterion for relaxing channels in Sec.~\ref{mixconvex} 
that will turn out to be extremely useful in our discussion. Our generalized setup gives rise to much richer quantum convergence effect - in Sect.~\ref{entanglement} we give numerical evidence for equilibrium
states which are entangled. Finally in Sect.~\ref{transport} we study the effects of having competitive
baths at different temperature.

\section{Mixing criteria in the homogenization setup}\label{mixconvex}

Our starting point is the observation that what  
makes the CP map in Eq.~(\ref{seq}) converge into
a specific state independently from the initial input $\rho_A$
is a well known property called {\em relaxing}~\cite{Terhal} (also referred to as ``mixing''~\cite{relaxing1,Giovannetti} or ``absorbing''\cite{relaxing2}).
In this language
the convergence point 
\begin{eqnarray}
\rho_A^* \equiv \lim_{n\rightarrow \infty} 
\rho^{(n)}_A \label{joj}\;,
\end{eqnarray}
 is the only {\em fixed point} of ${\cal E}$, i.e. 
the only solution of the equation
\begin{eqnarray}
{\cal E}(\rho_A) = \rho_A \;, \label{ffind}
\end{eqnarray}
(we refer
the reader to  Refs.~\cite{relaxing1,Giovannetti} for an  detailed introduction
to relaxing channels).
Therefore an homogenization procedure is associated with a 
relaxing channel whose fixed point $\omega_A$ coincides with the state of the 
controller.

We now prove a very simple but important
result. Suppose that given  $U$ we know
that the map~(\ref{mappa}) is relaxing for a specific choice
of the controlled state $\omega_B$. 
Consider now the map $\tilde{\cal E}$ which is obtained from Eq.~(\ref{mappa})
by replacing 
$\omega_B$ with a state $\tilde{\omega}_B$ which
is a (non trivial) convex combination of $\rho_B$, i.e. 
$\tilde{\omega}_B = p\; \omega_B + (1-p) \; \omega_B^\prime$,
with $p\in ]0,1]$ and $\omega_B^\prime$ being a generic density matrix.
In this case the map ${\tilde{\cal E}}$ can be expressed as
a  convex combination of ${\cal E}$, i.e.
\begin{eqnarray} \label{mixedmap}
\tilde{\cal E} = p \; {\cal E} + (1 -p)\; {\cal E}^\prime \;,
\end{eqnarray}
with the map ${\cal E}^\prime$ 
as in Eq.~(\ref{mappa}) with $\omega_B$ replaced by 
$\omega_B^\prime$. 
We can then use a theorem by Haag~\cite{Haag} 
which shows that a convex combination of CP maps
containing at least one relaxing channel is relaxing, 
to conclude that $\tilde{\cal E}$ is  relaxing
--- for completeness,
we provide an alternative (and much simpler) proof of this important
theorem. It is based on the fact relaxing map is equivalent to the asymptotic
deformation property~\cite{Giovannetti}. Hence for all $\rho\neq\rho'$
there is a $k$ such that 
\begin{eqnarray}
||{\cal E}^{k}(\rho)-{\cal E}^{k}(\rho^\prime)
||_{1}<||\rho-\rho^\prime||_{1} \;, \label{tracenorm}
\end{eqnarray} 
where
$\| \Theta\|_1 \equiv \mbox{Tr} [\sqrt{\Theta^\dag \Theta}]$ indicates 
 the trace norm of the operator $\Theta$.
We write $\tilde{\cal E}^{k}=p^{k}{\cal E}^{k}+(1-p^{k}){\cal S}^\prime,$ 
where ${\cal S}^\prime$
is the CP map that contains all other terms of the expansion of 
$\tilde{\cal E}^{k}.$
By the non-expansiveness~\cite{Ruskai} of ${\cal S}'$ and the triangle inequality
we obtain 
\begin{eqnarray}
||\tilde{\cal E}^{k}(\rho)-\tilde{\cal E}^{k}(\rho^\prime)|
|_{1}<||\rho-\rho^\prime||_{1}\label{nonexp} \;,
\end{eqnarray}
whence $\tilde{\cal E}$ is an asymptotic deformation. 
In the context of Fig.~\ref{figura2} this
implies that the system still converges when substituting the
controller state $\omega_B$ with $\tilde{\omega}_B$. For instance, if there
exists \emph{any} state $\omega_B$ for which the system converges, then it will converge
to the fully mixed state if the $B_k$ are initialized in the fully mixed state.

A natural question is then to determine the fixed point
 of  $\tilde{\cal E}$.
Specifically one may ask how
the final state of the system $A$  
depends upon the controller
state $\tilde{\omega}_B$. For instance: if homogenization takes place for $\omega_B$,
does it hold also for $\tilde{\omega}_B$?  Or, how does the entropy of the fixed point
depend on the entropy $S(\tilde{\omega}_{B})$ of the controllers? 

Before passing to apply the Haag criterion to the mediated homogenization 
scheme, it is worth presenting yet another interesting generalization of this 
simple but important theorem.
Consider in fact the situation in which 
the states of the controllers $B_1, \cdots, B_n$ have
not being properly initialized.
In particular we are interested in studying what happens
if instead of being prepared in the ``good'' initial state $\omega_B$, the $\ell$th controller 
is described by the following imperfect state 
\begin{eqnarray}
\bar{\omega}_{B}^{(\ell)} \equiv 
p_\ell \; \omega_{B} + (1-p_\ell) \; \varrho_{B}^{(\ell)}
\end{eqnarray}
where for $\ell =1, \cdots, n$,  $p_\ell>0$ 
are probabilities and $\varrho_{B}^{(\ell)}$ are density matrices.
According to the analysis of Sec.~\ref{s:theset} 
this yields a sequence ${\cal E}_{1} \cdots, {\cal E}_n$ of
CP maps which have the property that each of them is a convex combination
of a fixed relaxing map ${\cal E}$, i.e. 
\begin{eqnarray}
{\cal E}_{\ell}=
p_{\ell}{\cal E}+(1-p_{\ell}){\cal S}_{\ell}
\label{convex}\;,
\end{eqnarray}
where ${\cal E}$ and ${\cal S}_\ell$ are respectively the
channels~(\ref{mappa}) associated with $\omega_B$ and
$\varrho_B^{(\ell)}$ respectively. 
Clearly without putting any restriction on the values of $p_{\ell}$ nothing can be
said about the convergence property of the protocol.
Therefore we consider the case in which the ``error'' $(1-p_{\ell})$ is bounded, by imposing
the constraint
\begin{eqnarray}
p_{\ell}\geqslant p>0 \;.
\label{error}
\end{eqnarray}
This hypothesis does not yet guarantee that $A$
will be driven toward $\omega_A$. However we can 
at least 
verify that the process is still able to ``forget'' 
about the initial state of the
controlled system as in the relaxing case
(this is a typical feature of any homogenization protocol).
The evolution of the controlled system $A$  is in fact now 
described by the following sequence of concatenated maps,
\begin{eqnarray}
\mathcal{M}_{n}={\cal E}_{n}\circ {\cal E}_{n-1}
\circ\cdots\circ{\cal E}_{1} \label{forget}\;.
\end{eqnarray}
For arbitrary input density matrices $\rho'_A$, $\rho''_A$ 
of the controlled system define
\begin{eqnarray}
f_{n}=\|\mathcal{M}_{n}(\rho'_A)-\mathcal{M}_{n}(\rho''_A)
\|_1.
\label{efefe}
\end{eqnarray}
Now  since $f_n$ is non-negative and non-increasing~\cite{Ruskai} it certainly admits a limit 
$\lim_{n\rightarrow\infty}f_{n}\equiv f_{*}.$ 
To show that the protocol forces the controlled system
to forget about its initial conditions we need only to
verify that this quantity is null for all  $\rho'_A$ and
$\rho''_A$.
Assume then by contradiction that  $f_{*}>0$ for some choice
of these input states. Let $\rho^{*}_A$
be the fixed point of the unperturbed map ${\cal E}$. 
Then there is a value of $k$ such that
$\|\mathcal{E}^{k}(\rho_A)-\rho^{*}_A\|_{1}<f_{*}/4$ for \emph{all}
 $\rho_A$~\cite{Terhal}. Let then 
$\delta=\frac{p^{k}f_{*}}{3(1-p^{k})}>0.$
There is a $n$ such that $f_{n}-f_{*}<\delta.$ We write 
\begin{eqnarray}
\mathcal{M}_{k+n}=\tilde{\mathcal{M}}\circ\mathcal{M}_{n}
\end{eqnarray}
where the superoperator 
$\tilde{\mathcal{M}}={\cal E}_{k+n}\circ\cdots\circ
{\cal E}_{1+n}$
can be decomposed as 
\begin{eqnarray}
\tilde{\mathcal{M}}=P_{k}\mathcal{E}^{k}+(1-P_{k}) \Gamma
\label{fsfs}
\end{eqnarray}
with $\Gamma$ being CP and $P_{k}=p_{k+n}\cdots p_{1+n}\geqslant p^{k}$ by
assumption. Hence\begin{eqnarray*}
f_{k+n} & = & ||\tilde{\mathcal{M}}(\mathcal{M}_{n}(\rho_A))-
\tilde{\mathcal{M}}(\mathcal{M}_{n}(\rho'_A)||_{1}\\
 & \leqslant & P_{k}||\mathcal{E}^{k}(\mathcal{M}_{n}(\rho_A))
-\mathcal{E}^{k}(\mathcal{M}_{n}(\rho'_A))||_{1}\\
 &  & +(1-P_{k})||\Gamma(\mathcal{M}_{n}(\rho_A))-\Gamma
(\mathcal{M}_{n}(\rho_A))||_{1}\\
 & < & P_{k}f_{*}/2+(1-P_{k})(\delta+f_{*})\\
 & \leqslant & P_{k}f_{*}/2+(1-P_{k})\left[\frac{P_{k}f_{*}}{3(1-P_{k})}+f_{*}\right]\\
 & = & f_{*}-P_{k}f_{*}/6<f_{*}.\end{eqnarray*}
Since $f_{n}$ is non-increasing this is a contradiction, and $f_{*}=0.$
We have shown that the whole state space is contracted to a single
point. In general, this point is still evolving under the action of ${\cal E}_{n}$, but contains no information
about the initial state. The map $\mathcal{M}_n$ 
is relaxing if and only if there
exists an \emph{asymptotic} fixed point $\varrho^*_A$, 
i.e. a state with $\lim_{n\rightarrow \infty} \mathcal{M}_n(\varrho_A^*) = \varrho_A^*$. 

\section{Mediated Homogenization in spin networks}\label{s:heis}
An interesting example of mediated 
homogenization is obtained by assuming
$A$ to be a network of $N$ coupled qudits $A_1, \cdots, A_N$ 
mutually interacting through a sum of local 
term of the form 
\begin{eqnarray}
H_A=\sum_{k,k^\prime} J_{k k^\prime}S_{A_k A_{k^\prime}} \;, 
\label{hamilto}
\end{eqnarray}
where $J_{k k^\prime}$ are coupling constants  and where
$S_{A_k A_{k^\prime}} = (
S_{A_k A_{k^\prime}})^\dag$ are unitary operators 
which swap the $k$-th qudits of $A$ with 
the $k^\prime$-th~\cite{NOTAswap}. 
Regarding the coupling with the controller we consider the case 
in which only the $A_N$ 
interact with the 
$B$s (also represented by $d$-dimensional systems) 
through a swap Hamiltonian similar to~(\ref{hamilto}), i.e.
\begin{eqnarray}
H_I= S_{B A_N} 
\label{ddef}
\;.
 \end{eqnarray}
Under these conditions we can show  that, for all choice of the
controller states $\omega_B$ and for almost all choices of the
interaction time $t$ the map~(\ref{mappa}) is relaxing with fixed point
\begin{eqnarray}
\rho_A^* = \left(\omega_A\right)^{\otimes N}\;,
\label{fix1}
\end{eqnarray}
given that the graph associated with the coupling $J_{kk^\prime}$ satisfies
certain constraints.
This corresponds to the case in which, in the limit of large $n$,
the controller state $\omega_B$ 
is ``copied''  in  all the $N$ controlled  qudits. We call this process
 a {\em mediated homogenization} of $A$. It fulfills all four homogenization criteria
mentioned in Ref.~\cite{Scarani}: Firstly, the coupling between system and bath is independent of the bath state. Secondly, the equilibrium state is not only a fixed point of the CP map~(\ref{mappa}) but also of the unitary evolution$U\equiv {\exp}[ -i (H_A + H_I)t ]$ alone. Thirdly, the system converges to the fixed point for all initial states. Finally the change of the bath due to the evolution can be made arbitrarily small by choosing a short interaction time $t$. 
An immediate consequence of the above result is the fact that
the von Neumann entropy of $A$  converges to $N$ times the von Neumann
entropy of the controller state $S_{B} = - \mbox{Tr} [ \rho_B \log_2 
\rho_B]$. This is a distinctive  trait of the mediated homogenization processes and it is 
similar to what happens when we put a system of interest in thermal
contact with an reservoir. It should be pointed out though that in our case
 the convergence state is in general far away from any thermal state $\exp [-\beta H_{A}]/Z$. The thermalization
feature observed in~\cite{HOMOGENIZATION1} thus seems to be specific to the case where $A$ is a single qudit.

To prove the above result we first focus on the case in which 
$\omega_B$ is a pure vector $|\phi\rangle_B$.
Define  then the joint observable
\begin{eqnarray}
M_{AB} = M_B + M_A \label{obse} \;,
\end{eqnarray} 
where $M_A = \sum_{k=1}^N M_{A_k}$ and 
\begin{eqnarray}
M_{B} &\equiv& - |\phi\rangle_B \langle \phi| \nonumber \\
M_{A_k} &\equiv& - |\phi\rangle_{A_k}\langle \phi|  \;.
\end{eqnarray}
The operator $M_{AB}$ commutes
with the total Hamiltonian $H = H_A + H_{BC}$ and hence with 
the operator $U= \exp[-i Ht]$: we thus say that free
evolution of the network preserves the ``excitations'' associated with
the projectors $|\phi\rangle\langle \phi|$.

Moreover the state 
$|\phi\rangle_B$ is the (non degenerate) 
eigenvector associated with the minimum eigenvalue of $M_B$.
Under these conditions we can invoke the Lemma 3  of Ref.~\cite{Giovannetti}
which states that the map~(\ref{mappa}) is relaxing with fixed point
$|\phi\rangle^{\otimes N} = |\phi\rangle_{A_1} \otimes \cdots \otimes
 |\phi\rangle_{A_N}$ if the state $|\phi\rangle^{\otimes N} \otimes
|\phi\rangle_B$   is the unique eigenvector of $U$
having the form
$|E\rangle_{A} \otimes |\phi\rangle_B$.  
This last condition can be verified by focusing on the  global Hamiltonian $H$:
 if  indeed  $|\phi\rangle^{\otimes N} \otimes |\phi\rangle_B$
is the unique  eigenvector
of the global Hamiltonian $H$  with the factorization property $|E\rangle_A
\otimes |\phi\rangle_B$ then the same property will holds for $U$ for almost
all the values of $t$.
We have shown  elsewhere~\cite{Burgarth2007} that for ``excitation'' preserving
Hamiltonians 
the above factorization condition 
depends  only on the geometry of
 the associated graph. For example, an open chain with $C$
being an end qudit has the required property. 
We can apply this result to the Hamiltonian~(\ref{hamilto}) (
it is excitations preserving in the sense that it does preserve the
number of qudits of $A$ which are in the state $|\phi\rangle$).
Therefore for any given network configuration satisfying the 
topological constraint of Ref.~\cite{Burgarth2007} we can conclude that,
for all initial pure state $|\phi\rangle$ of the controller,
the above iterative procedure will drive $A$ to a unique fixed point.
Before determining such fixed point, let first observe that
the same convergence will hold also when assuming the initial state of the
controllers to be a general mixed state $\omega_B$.
This is a trivial consequence of the pure case scenario 
which can be obtained by expanding any such mixture into
a convolution of pure states $\omega_B = \sum_j q_j |\phi_j\rangle_B\langle
\phi_j|$ and applying the Haag criterion. 

Since we have now proved that for all choice of the controller
state $\omega_B$ the  channel ${\cal E}$ is relaxing, to  verify
Eq.~(\ref{fix1}) it is sufficient to show that $\left(\omega_A\right)^{\otimes N}$
satisfies Eq.~(\ref{ffind}). 
The latter can be easily verified by noticing that 
each summand of the Hamiltonian~(\ref{hamilto}) 
commutes with all tensor product operators
of the form $\Theta^{\otimes N}$, and therefore
\begin{eqnarray}
[ H_A, \Theta^{\otimes N}  ] = 0 \label{cc} \;.
\end{eqnarray}
Consequently 
for $\rho_A^{*}$ as in Eq.~(\ref{fix1}) and $H_{BC}$ as
in Eq.~(\ref{ddef}) we 
 can write
\begin{eqnarray}
[ H_A + H_{BC} , \rho_A^{*} \otimes \rho_B ] = 0  
\Longrightarrow 
\label{eed}
[ U , \rho_A^{*} \otimes \rho_B ] = 0  \;,
\label{eed1}
\end{eqnarray}
which is sufficient to show that $\rho_A^*$ satisfies the 
invariance condition~(\ref{ffind}).

\section{Building entanglement in the network}\label{entanglement}
In the previous section we found a model
where independently from the initial
state of $A$, the final state of the network is the \emph{separable} state
$\omega_A^{\otimes N}$.
Each of the $N$ qudits of the network 
has been driven into the initial state of the controllers.
In this section
we show 
that, keeping  $H_I$ as in Eq.~(\ref{ddef}),
 there are also Hamiltonians $H_A$ 
which are capable of building entanglement among the qudits of the network.
Although this is no longer a homogenization protocol (the controller state
is not transferred to the controlled system) it could have useful applications
as a method of state preparation.

\begin{figure}[ht]
\includegraphics[width=0.8\columnwidth]{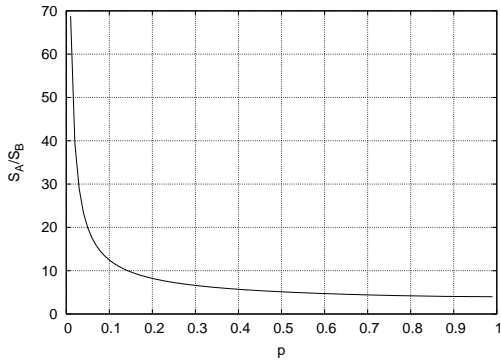}
\caption{\label{FXY}Ratio $R$ of the {}``output'' entropy $S_{A}$ of the convergence
point and the {}``input'' entropy $S_{B}$ of the bath state for
a XX chain of length $N=4.$
The controller state is as in Eq.~(\ref{example1}).
For all values of $p$ the dynamics is relaxing --- for $p>0$ this
is a trivial consequence of Haag theorem, for $p=0$ instead it
can be directly proved by numerical means~\cite{NUMERICS}. When $p\rightarrow 1$ the
state becomes diagonal and the ratio converges to $N$. For 
$p\rightarrow0$ the ratio diverges as the bath state 
becomes pure but the convergence point
remains mixed. 
This should be compared with the behavior of a mediated
homogenization process (e.g. the swap coupling of Sec.~\ref{s:heis})
where
$R$  is always equal to 
$N$ for all $p$. The parameters for the numerics are $J_{kk^\prime}=\delta_{k,k+1}$ and $t=0.5$.}
\end{figure}

Consider for the sake of simplicity $d=2$.
In this case
the swap interaction of Eq.~(\ref{hamilto}) corresponds (up to a constant) 
 to a Heisenberg coupling. 
A natural generalization of it
is then provided by
the anisotropic Heisenberg  Hamiltonian
\begin{eqnarray} 
H_A= \sum_{k,k^\prime}  \frac{J_{k k^\prime}}{2} \left(
{\sigma}^{(x)}_{k}{\sigma}^{(x)}_{k^\prime}
+{\sigma}^{(y)}_{k}{\sigma}^{(y)}_{k^\prime}+ \Delta\;  
{\sigma}^{(z)}_{k}{\sigma}^{(z)}_{k^\prime}
\right),
\label{hamilto1}
\end{eqnarray}
where $\sigma_{k}^{(x,y,z)}$ represents the Pauli matrix of the $k$-th
qubit of $A$ and where $\Delta -1$ is the anisotropy parameter
(the isotropic
coupling is obtained for $\Delta =1$). 
For this coupling we can use the same argument given in previous section
to characterize the relaxing properties of the associated map~(\ref{mappa}) 
(in particular the
factorization
property of its eigenvectors
depends only on the geometry of the associated graph~\cite{Burgarth2007}). 
In this case however 
the isotropy is lost and the Hamiltonian has a preferred spatial direction 
associated with the $\hat{z}$ axis which makes $|0\rangle_B$ and 
$|1\rangle_B$ special with respect to the other controller pure states. 
Indeed   we  can still show that mediated homogenization takes place for 
input states $\omega_B$ which are diagonal in the computational basis, i.e.
\begin{eqnarray}
\omega_{B}=p|0\rangle_{B}\langle0|+(1-p)|1\rangle_{B}\langle1| \;.
\end{eqnarray}
This follows by the fact that 
 for such a choice Eq.~(\ref{eed1}) holds independently from the
value of $\Delta$. 
On the contrary for more general controller states 
mediated homogenization is lost.
As an example,
consider  
\begin{eqnarray}
\omega_{B}=p|0\rangle_{B}\langle0|+(1-p) 
|-\rangle_{B}\langle -|
\label{example1}
\end{eqnarray}
 where $| -\rangle_B \equiv 
\left(|0\rangle_{B}- |
1\rangle_{B}\right)/\sqrt{2}$ and $p>0$.
In this case  Haag's theorem can still be used to ensure
relaxing of the map~(\ref{mappa}) even though  computing the fixed
point is not simple.  For such choice  however we have numerically
verified that the mediated homogenization  does not take place in general.
We  evaluated 
the asymptotic limit of the von Neumann entropy 
of $\rho_A^{*}$,  verifying that it is no longer
 a multiple of the von Neumann entropy of
the bath state.
In Fig.~\ref{FXY} we show an example for a
$XX$ chain ($\Delta =0$). 

Of particular interest is the case $p=0$.  
In this limit $\omega_B = |-\rangle_B\langle -|$ and
the relaxing property cannot be established 
from the theorem by Haag (simply  $\rho_B$ is not 
a convex combination of $|0\rangle_B\langle 0|$). Nevertheless we 
can use numerical analysis to show that the map (\ref{mappa}) is still relaxing~\cite{NUMERICS}.  
We found that the convergence point is highly mixed.
%%%%%%%%%%%%%%%%%%%%%%%%%%%%%%%%%%%%%%%
\begin{figure}[ht]
\includegraphics[width=0.8\columnwidth]{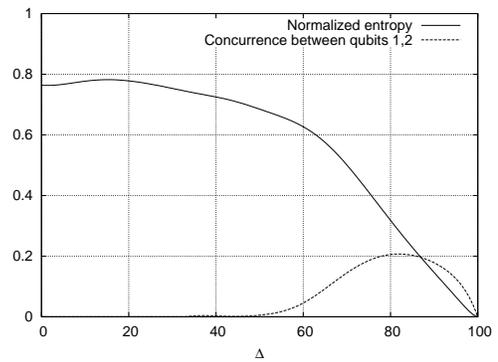}
\caption{\label{FXYD}The entropy $S_{A}$ of the convergence point and the concurrence between the first two qubits for an anisotropic Heisenberg chain
of length $N=4$ as a function of the anisotropy parameter $\Delta.$ The bath state is given
by $(|0\rangle_{B}-|1\rangle_{B})/\sqrt{2}.$ The parameters for the numerics are $J_{kk^\prime}=\delta_{k,k+1}$ and $t=0.5$. }
\end{figure}
 Since this is so much different from the isotropic Heisenberg model with fixed point $|-\rangle^{\otimes N}$ it seemed natural to  compute the relaxing property and convergence point of the
anisotropic model for $p=0$ as a function of $\Delta$ to see the transition
for a $XX$ chain (with highly mixed convergence point) to the Heisenberg
chain (pure convergence point). In particular we wanted to check if
there are also entangled fixed points. For this purpose we computed
the concurrence between the first and second qubit of the chain (say)
for intermediate $\Delta$ (see Fig.~\ref{FXYD}). Again, for the
given parameters, all examples were relaxing. We found that the convergence point is indeed entangled for some
values of $\Delta$. Contrary to the results in the last section, the numerical examples of convergence points observed here depend on the parameters of the model.
An important open problem  is to determine if there exist  $H_A$ and $\omega_B$
that have a fixed point with interesting applications (e.g. a cluster state).

\section{Dynamical equilibrium}\label{transport}

\begin{figure}[t]
\includegraphics[width=0.7\columnwidth]{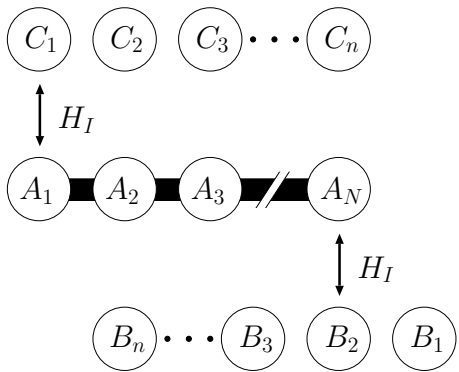}
\caption{\label{duo} Setup of the dynamical equilibrium: here the system of Fig.~\ref{figura2} is coupled
to two competing baths at different temperatures.}
\end{figure}

\begin{figure}[t]
\includegraphics[width=0.8\columnwidth]{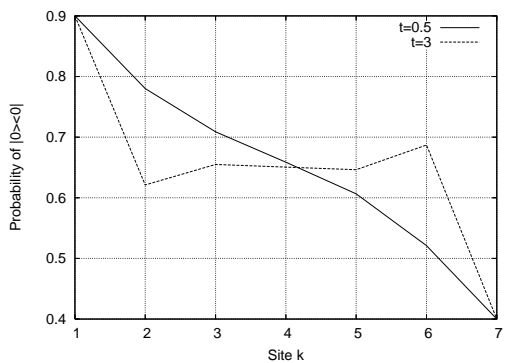}
\caption{\label{F1} Probability of finding the $k$th qubit in the state $|0\rangle\langle 0|$ for the steady state of a Heisenberg spin chain of length $N=5$. The plot includes the systems $B$ (site $1$)  and $C$ (site $7$).The parameters $p=0.9$ and
$q=0.4.$ We give two examples with different choice of the interaction time $t$.}
\end{figure}

The many-body structure of $A$ presented in Fig.~\ref{figura2} 
allows us to consider more complicated procedures. For instance 
we can analyze  \emph{competitive} configurations where the dynamics of the network $A$ is driven by the 
simultaneous coupling with two independent sets of controllers
(the $B_1,\cdots, B_n$ and the $C_1, \cdots, C_n$ of Fig.~\ref{duo}). 
%%%%%%%%%%%%%%%%%%%%%%%%%%%%%%%%%%%%%%%%%%%%%%%%%%%%
%%%%%%%%%%%%%%%%%%%%%%%%%%%%%%%%%%%%%%%%%%%%%%%%%%%%%%%%%%
We can then model the ``transport'' of excitations
through the network by assuming the two sets to be directly coupled
with distinct network elements (say  $A_N$ for  $B$ and $A_1$ for $C$)
and assuming different ``temperature'' for the two species of controllers
(say 
$\omega_{B}=p|0\rangle_{B}\langle0|+(1-p)|1\rangle_{B}\langle1 |$
for the $B$s
and $\nu_C = q|0\rangle_{C}\langle0|+(1-q)|1\rangle_{C}\langle1|$
for the $C$s).
A similar situation is considered in~\cite{Michel2003,b2, Mejia-Monasterio2005}
where in the case of a linear chain coupled through Heisenberg and XX interactions   
the relaxing property was observed \emph{numerically}~\cite{NOTE}.  Here, the convergence can be derived analytically for arbitrary chain length 
as a consequence of the Haag theorem.
To verify this 
it is convenient  to treat  $B$ and $C$ as a unique
controller composed by elements $B_1 C_1, B_2 C_2, \cdots, B_n C_n$.
From the above definitions it then follows that such composite controllers
are initialized in the state 
\begin{eqnarray}
\omega_B\otimes \nu_C = 
pq \; |0\rangle_{B}\langle 0|\otimes |0\rangle_C\langle 0| +(1-pq)\; \varrho_{BC}
\label{afafa}\;,
\end{eqnarray}
with $\varrho_{BC}$ being a density matrix. 
Therefore according to the Haag theorem the convergence  
can be verified by focusing only on the case in which
$B$ and $C$ are initialized in 
$|0\rangle_{B}\langle 0|\otimes |0\rangle_C\langle 0|$. 
With this choice however the iterative procedure is equivalent to the
``cooling'' protocol discussed in Ref.~\cite{Burgarth2007} 
and the convergence is automatically verified.
Deriving
the exact steady state in this case is however quite complicated 
so we restrict to numerical analysis. Again its form depends strongly on the parameters, as shown in Fig.~\ref{F1}.
%%%%%%%%%%%%%%%%%%
%%%%%%%%%%%%%%%%%%%%%%%%%%%%

\section{Conclusion}

We have generalized the homogenization protocols to a scenario where the
system is no longer a single qudit. We found that mediated homogenization
still takes place on the lattice when the interaction is taken to be isotropic. Anisotropic interactions on the other hand do not in general show homogenization. 
Our numerical results are quite suggestive in this direction but are certainly not
conclusive.
This suggests many further studies: what is the structure of the fixed points of these systems? How are their entropies related to the bath entropy? What happens when the system is close to a critical point? Can we use this convergence as a way of preparing \emph{useful} states such as cluster states on optical lattices?  Finally we looked at transport along chains interconnecting baths at different temperature, where the Haag criterion allowed us to prove the convergence to a dynamical equilibrium. We found that the temperature profile is strongly depending on the parameters of the system, such as the interaction time, and not even monotonic for some times. While at the moment these results are numerically only, it may be possible to obtain an analytic expression for the fixed point in a weak coupling limit by deriving a closed equation for the proper ansatz (cf.~\cite{b2}). This will be subject of future investigations.

DB would like to thank Mario Ziman, Vladimir Buzek, Valerio Scarani, Nicolas Gisin and Heinz-Peter Breuer for many fruitful discussions. He acknowledges financial
support by the Swiss National Science Foundation (SNSF).


\begin{thebibliography}{8}
\bibitem{NOTAswap}
The swap operator is defined by $S_{A_k A_{k^\prime}} |\psi\rangle_{A_k}\otimes |\phi\rangle_{A_k\prime}=|\phi\rangle_{A_k}\otimes |\psi\rangle_{A_k\prime}$ for all  $|\psi\rangle_{A_k}, |\phi\rangle_{A_k\prime}$. It is hermitian and unitary.
\bibitem{NUMERICS}The method we chose to verify this is to calculate
the spectrum of the map ${\cal E}$ and show that there is exactly
one eigenoperator with maximal eigenvalue $|\lambda|=1.$ As discussed 
in~\cite{Terhal,Giovannetti} this is in fact 
sufficient to show that the channel ${\cal E}$ is 
relaxing.
\end{thebibliography}
\end{document}